\documentclass[11pt,oneside,letter]{article}

\usepackage{geometry}
\usepackage{amsmath, amsthm,pdfpages,url}
\usepackage{amsfonts,amssymb,dsfont}
\usepackage{nicefrac,mathrsfs}
\usepackage{bm}                                 %
\usepackage[backgroundcolor=white,bordercolor=orange]{todonotes}
\usepackage{bbm}

\usepackage{booktabs} 
\usepackage{multirow}
\usepackage{soul} 
\usepackage{microtype}
\usepackage{upgreek,adjustbox,multirow,makecell,diagbox} 

\usepackage{natbib}

\newtheorem{theorem}{Theorem}
\newtheorem{proposition}[theorem]{Proposition}

\theoremstyle{definition}

\theoremstyle{remark}
\newtheorem{remark}[theorem]{Remark}

\renewcommand{\P}{\mathbb{P}}

\newcommand{\Ind}{{\mathds 1}}

\newcommand{\E}{\mathbb{E}}

\renewcommand{\P}{\mathcal{P}}

\newcommand{\beq}{\begin{equation}}
\newcommand{\eeq}{\end{equation}}

\renewcommand{\P}{\PP}
\newcommand{\Q}{\mathbb{Q}}

\newcommand{\dbra}[1]{[\kern-0.15em[ #1 ]\kern-0.15em]}

\newcommand{\ind}[1]{\Ind_{\{#1\}}}

\newcommand{\ccN}{{\mathscr N}}\newcommand{\cN}{{\mathcal N}}

\newcommand{\heute}{\number\day. \ifcase\month\or
  Januar\or Februar\or Maerz\or April\or Mai\or Juni\or
  Juli\or August\or September\or Oktober\or November\or Dezember\fi
  \space \number\year}
\newcommand{\dbraco}[1]{[\kern-0.15em[ #1 [\kern-0.15em[}
\newcommand{\smalltonormalsize}{%
  \fontsize
  {\fpeval{(\f@size@small+\f@size@normalsize)/2}}
  {\fpeval{(\f@baselineskip@small+\f@baselineskip@normalsize)/2}}%
  \selectfont
}

\renewcommand{\P}{\mathbb{P}}

\usepackage{eurosym}
\DeclareRobustCommand{\officialeuro}{%
	\ifmmode\expandafter\text\fi
	{\fontencoding{U}\fontfamily{eurosym}\selectfont e}}

\usepackage{comment}
 
\usepackage{overpic}

\usepackage[colorlinks=true, pdfstartview=FitV, linkcolor=blue,
            citecolor=blue, urlcolor=blue]{hyperref}

\title{Machine learning for  multiple yield curve markets: fast calibration in the Gaussian affine framework}

\author{Sandrine G\"umbel  and Thorsten Schmidt\thanks{Department of Mathematical Stochastics, University of Freiburg, Ernst-Zermelo Str. 1, 79104 Freiburg im Breisgau, Germany \newline
\hspace*{1.45em}    Emails: \url{sandrine.guembel@stochastik.uni-freiburg.de}, \url{thorsten.schmidt@stochastik.uni-freiburg.de}\newline
Financial support the German Research Foundation (DFG) within project No. SCHM 2160/9-1  is gratefully acknowledged.}
 }
\begin{document}

\maketitle

\begin{abstract}
Calibration is a highly challenging task, in particular in multiple yield curve markets. This paper is a first attempt
to study the chances and challenges of the application of machine learning techniques for this. 
We employ Gaussian process regression, a machine learning methodology having many similarities with extended K\'alm\'an filtering - a technique which has been applied many times to interest rate markets and term structure models. \\
We find  very good results for  the single
 curve markets and many challenges for the multi curve markets in a Vasi\v cek framework. The Gaussian process regression is implemented with the Adam optimizer and the non-linear conjugate gradient method, where the latter performs best.  We also point towards future research. 
\end{abstract}


\section{Introduction}

It is the aim of this paper to apply machine learning techniques to the calibration of bond prices in multi curve markets in order to predict the term structure of basic instruments. The challenges are two-fold: on the one side interest rate markets are characterized by having not only single instruments, like stocks, but full term structures, i.e. the curve of yields for different investment periods. On the other side, in multi curve markets not only one term structure is present but multiple yield curves for different lengths of the future investment period are given in the market and have to be calibrated. This is a very challenging task, see \cite{Eberlein2018} for an example using L\'evy processes.  

The co-existence of different yield curves associated to different tenors is a phenomenon in interest rate markets which originates with the 2007--2009 financial crisis. In this time,  spreads between different yield curves reached their peak beyond 200 basis points. Since then the spreads have remained on a non-negligible level.  The most important curves to be considered in the current economic environment are the overnight indexed swap (OIS) rates and the interbank offered rates (abbreviated as Ibor, such as Libor rates from  the London interbank market) of various tenors. In the European market these are respectively the Eonia-based OIS rates and the Euribor rates. The literature on multiple yield curves is manifold and we refer to \cite{GrbacRunggaldier} and \cite{Henrard2014} for an overview. The general theory and affine models have been developed and applied, among others, in \cite{Mercurio2010,Grbac2015affine,Cuchiero2016,Cuchiero2016affine,Grbacea2020}.     

The recent developments have seen many machine learning techniques, in particular deep learning became very popular. While deep learning typically needs big data, here we are more confronted with \emph{small data} together with a high-dimensional prediction problem, since a full curve (the term structure), and in the multi curve market, even multiple curves have to be calibrated and predicted. To be able to deal with this efficiently, one would like to incorporate information from the past, and a Bayesian approach seems best suited to this. 
We choose {\em Gaussian process regression} (GPR)\index{Gaussian process regression} as our machine learning approach which ensures fast calibration; see \cite{de2018machine}. This is a non-parametric Bayesian approach to regression and is able to capture non-linear relationships between variables.  
The task of {\em learning} in Gaussian processes simplifies to determining suitable properties for the covariance and mean function, which will determine the calibration of our model. We place ourselves in the context of the Vasi\v cek model, which is a famous affine model, see \cite{Filipovic2009} and \cite{KellerResselSchmidtWardenga2018} for a guide to the literature and details. %

\subsection{Related literature}
Calibration of log-bond prices in the simple Vasi\v cek model framework via Gaussian processes for machine learning has already been applied in \cite{Beleza2012} for a single maturity and in \cite{Beleza2014} for several maturities. They both rely on the theory of Gaussian processes presented in \cite{Rasmussen2006}. While in Section~\ref{subsec:singleVas}
we will extend \cite{Beleza2012} by presenting an additional optimization method, \cite{Beleza2014}  and Section~\ref{subsec:multiVasicek}  constitute a different access to the calibration of interest rate markets with several maturities. While \cite{Beleza2014} calibrate solely zero-coupon log-bond prices, from which one is not able to construct a post-crisis multi-curve interest rate market, we calibrate zero-coupon log-bond prices and log-$\delta$-bond prices on top in order to encompass forward rate agreements and to be conform with the  multi-curve framework, cf. \cite{Grbacea2020} for the notion of $\delta$-bonds. The modelling of log-$\delta$-bond prices allows to build forward-rate-agreements (FRAs) so that we have the basic building blocks for multi-curve interest rate markets. 

\section{Gaussian process regression}

Following \cite[Chapter 2 and 5]{Rasmussen2006} we provide a brief introduction to Gaussian process regression (GPR). For a moment consider a  regression problem with  additive Gaussian noise: we observe $y_1,\dots,y_n$ together with covariates $x_1,\dots,x_n$ and assume that
$$ y_i = f(x_i,\theta) + \epsilon_i, \qquad i=1,\dots,n,$$
where $f(\cdot,\theta)$ is a regression function depending on the unknown parameter $\theta$ and $\epsilon$ is a $d$-dimensional noise vector which we assume to consist of i.i.d.~mean-zero Gaussian errors and standard deviation $\hat\sigma$. 

In the Bayesian approach we are not left without any knowledge on $\theta$ but may start from a \emph{prior} distribution; sometimes this distribution can be deducted from previous experience in similar experiments, while otherwise one chooses an uninformative prior. Assuming continuity of the prior, we  denote the prior density of $\theta$ by $p(\theta)$. Inference is now performed by computing the 
 \emph{a posteriori} distribution of $\theta$ conditional on the observation $(x,y)$. This can be achieved by Bayes' rule, i.e.~
$$ p(\theta |x,y) = \frac{p(\theta,y | x)}{p(y|x)} = \frac{ p(y|x,\theta) p(\theta)}{p(y|x)}, $$
where we assumed only that the distribution of $\theta$ does not depend on $x$. Similarly, we can compute $p(y|x)$ from $p(y|x,\theta)$ by integrating with respect to $p(\theta)$.

For a moment, we drop the dependence on $x$ in the notation.  Assuming only that the observation $y\sim \ccN(\mu_y,\Sigma_{yy})$ is normally distributed, 
we are already able to state the \emph{marginal likelihood} $p(y)$: it is given up to a normalizing constant $c$ by
\begin{align}\label{eq:logmarglik}
  \log p(y) = c - \frac 1 2 (y-\mu_y)^\top \Sigma_{yy}^{-1} (y- \mu_y).
\end{align}

If we assume moreover that $\xi=f(\theta)$ is jointly normally distributed with $y$, we arrive at the multivariate Gaussian case. Hence, the conditional distribution $p(\xi|x,y)$ is again Gaussian and can be computed explicitly. 
Starting from $(\xi,y)^\top \sim \cN(\mu,\Sigma)$, where we split $\mu=(\mu_\xi,\mu_y)^\top$ and 
$$ \Sigma = \left(\begin{matrix} \Sigma_{\xi \xi} & \Sigma_{\xi y} \\
\Sigma_{y\xi} & \Sigma_{yy} \end{matrix}\right), $$
we can compute $p(\xi|y)$ through some  straightforward calculations\footnote{We give a short derivation in the Appendix.}. First, 
observe that $y \sim \ccN( \mu_y, \Sigma_{\xi \xi} + \hat\sigma^2 I_n)$. We  obtain $\Sigma_{yy} = \Sigma_{\xi \xi} + \hat\sigma^2 I_n $. Hence
the \emph{marginal likelihood} is given by
\begin{align}\label{eq:marglik}
  \log p(y) = \tilde c - \frac 1 2 (y-\mu_y)^\top \big( \Sigma_{\xi \xi} + \hat\sigma^2 I_n \big)^{-1} (y- \mu_y).
\end{align}

Second, we compute
\begin{align} \label{updating}
\xi | y \sim \ccN\Big( \mu_\xi + \Sigma_{\xi y} \Sigma_{yy}^{-1}(y-\mu_y), \Sigma_{\xi \xi}- \Sigma_{\xi y} \Sigma_{yy}^{-1}\Sigma_{y\xi}\Big).
\end{align}
This formula is the basis for the calibration in the Vasi\v cek model, as we will show now.

\section{The single-curve Vasi\v cek interest rate model}\label{subsec:singleVas}

As a first step, we calibrate the single-curve Vasi\v cek interest rate model following \cite{Beleza2012}. To begin with, we want to mention the important difference in mathematical finance between the objective measure $\P$ and the risk-neutral measure $\Q$. The statistical propagation of all stochastic processes takes place under $\P$. Arbitrage-free pricing means computing prices for options; of course the prices depend on the driving stochastic factors. The fundamental theorem of asset pricing now yields that arbitrage-free prices of traded assets can be computed by taking expectations under a risk-neutral measure $\Q$ of the discounted pay-offs. For a \emph{calibration}, the risk-neutral measure has to be fitted to observed option prices, which is our main target.

In this sense, we consider zero-coupon bond prices under the risk-neutral measure $\Q$.
The Vasi\v cek model is a  single-factor model  driven by the short rate $r=(r_t)_{t \ge 0}$ which is given by the solution of the stochastic differential equation
\begin{equation}\label{eq:r}
  dr_t = \kappa (\theta - r_t )dt + \sigma dW_t, \quad t \ge 0,
\end{equation}
with initial value $r_0$. Here  $W$ is a $\Q$-Brownian motion and $\kappa, \theta, \sigma$ are positive constants. For a positive $\kappa$ the process $r$ converges to the long-term mean $\theta$. 

The price process of the  zero-coupon bond price with maturity $T$ is denoted by $(P(t,T))_{0 \le t \le T}$ with $P(T,T)=1$. The Vasi\v cek model is an affine bond price model, which implies that bond prices take an exponential affine form, i.e.
\begin{equation}\label{eq:Vasicek_recall}
P(t,T) = e^{-A(T-t)-B(T-t) r_t} \qquad \text{for } t \leq T.
\end{equation}
Here, 
the functions $(A(\cdot),B(\cdot)):[0,T] \rightarrow \mathbb{R} \times \mathbb{R}$ are given by
\begin{align}\label{eq:Ricrecall}
A(T-t) &=  \frac{\theta}{\kappa } \left(e^{-\kappa (T-t) } + \kappa (T-t) -1\right)\nonumber\\
&\quad +\frac{ \sigma ^2}{4 \kappa ^3}\left(e^{-2 \kappa (T-t) }-4 e^{- \kappa(T-t) }-2
\kappa(T-t) +3 \right),\nonumber\\
B(T-t) &=  \frac{1}{\kappa }(1 - e^{-\kappa(T-t)  }).
\end{align}

We recall that the solution of the stochastic differential equation \eqref{eq:r} is given by (cf. \cite[Chapter 4.2]{BrigoMercurio01}), 
\begin{align*}
  r_t = r_0 e^{-\kappa t} + \theta (1-e^{-\kappa t}) + \sigma e^{-\kappa t} \int_0^t e^{\kappa u} dW_u.
\end{align*}

Together with \eqref{eq:Vasicek_recall}, this implies that the zero coupon bond prices are log-normally distributed. To apply the Gaussian process regression,  we consider in the sequel log-bond prices for $t \leq T$,
\begin{equation*}
  y(t,T) = \log P(t,T) = -A(T-t)-B(T-t) r_t.
\end{equation*}
The  corresponding mean function is given by
\begin{align}\label{eq:mean0vas}
  \mu(t,T)& := \E_{\Q}\big[\log P(t,T)\big]
= -A(T-t)-B(T-t)\big(r_0 e^{-\kappa t} + \theta (1-e^{-\kappa t})\big),
\end{align}
and  the covariance function is given by
\begin{align}\label{eq:cov0vas}
  c (s,t,T) &:= \E_{\Q}\Big[\big(\log P(s,T) - \E_{\Q}[\log P(s,T)]\big)\cdot\big(\log P(t,T) - \E_{\Q}[\log P(t,T)]\big)\Big]\nonumber\\
  &= B(T -s)B(T -t)\Big(\E_{\Q}[r_{s} r_{t}] - \E_{\Q}[r_{s}]\E_{\Q}[r_{t}]\Big)\nonumber\\
  &= B(T -s)B(T-t) \frac{\sigma^2}{2\kappa} e^{-\kappa(s +t)} \big(e^{2\kappa (s \wedge t)} -1\big).
\end{align}
An \emph{observation} now consists of  a vector
$y=(y(t_1,T),\dots,y(t_n,T))^\top + \epsilon$ of log-bond prices with additional Gaussian i.i.d.~noise with variance $\hat\sigma^2$. Consequently,  
 $y \sim \ccN(\mu_y,\Sigma_{yy} ) $ with
\begin{align}\begin{aligned} 
  \mu_y &= (\mu(t_1,T),\dots,\mu(t_n,T))^\top\\
  (\Sigma_{yy})_{i,j} &= c(t_i,t_j,T) + \hat\sigma^2 \ind{i = j}. \end{aligned} \label{distribution y}
\end{align}
We define the vector of hyper-parameters as $\Theta = (r_0, \kappa, \theta, \sigma)^\top$ and aim at finding the most reasonable values of hyper-parameters, minimizing the negative log marginal likelihood defined in \eqref{eq:logmarglik} with the underlying mean  covariance function from \eqref{distribution y}.

\begin{remark}[Extending the observation] Note that in the set-up considered here, the driving single factor process $r$ is  considered not observable at $t_0,t_1,\dots,t_n$, compare Equation \eqref{distribution y} which depends on $r_0$ only (and on $\Theta$ of course). Such an approach  may be  appropriate for short time intervals. The propagation of $r_0$ to the future times $t_1,\dots,t_n$ is solely based on the risk-neutral evolution under the risk-neutral measure $\Q$ via $\mu$ and $c$. 

If, on the contrary, updating of $r$ should be incorporated, the situation gets more complicated since the evolution of $r$ (taking place under the statistical measure $\P$) needs to be taken into account. 
One can pragmatically proceed as follows: assume that at each time point $r$ is observable. We neglect the information contained in this observation about the hyper-parameters $\Theta$. Then, the formulas need to be modified only slightly by using conditional expressions. For example, $\mu(t,T)$ conditional on $r_s$ for $s \le t$ equals
$$ \E_\Q[\log P(t,T)|r_s] = -A(T-t)-B(T-t)\big(r_s e^{-\kappa (t-s)} + \theta (1-e^{-\kappa (t-s)})\big), $$
and, if $t=s$, this simplifies even further. 

In a similar spirit, one may extend the observation by adding bond prices with different maturities or,  additionally, derivatives. A classical calibration example would be the fitting of the hyper-parameters to the observation of bond prices and derivatives, both with different maturities, at a fixed time $t_0$. The Gaussian process regression hence also provides a suitable approach for this task.
We refer to \cite{de2018machine} for an extensive calibration example in this regard.
\end{remark}

 We refer to \cite{Rasmussen2006} for other approaches to model selection such as cross validation or alignment. For our application purposes maximizing the log-marginal likelihood is a good choice since we already have information about the choice of covariance structure, and it only remains to optimize the hyper-parameters, cf. \cite{Fischer2016}.

\subsection{Prediction with Gaussian Processes regression}
The prediction of new log-bond prices given some training data is of interest in the context of the risk management of portfolios and options on zero-coupon bonds and other interest derivatives. The application to pricing can be done along the lines of \cite{DuembgenRogers}, which we do not deepen here.
Once having found the  calibrated parameters $\Theta^*$, we can apply those calibrated parameters to make predictions for new input data. 

\begin{figure}[!t]
  \centering
  \includegraphics[width=\textwidth]{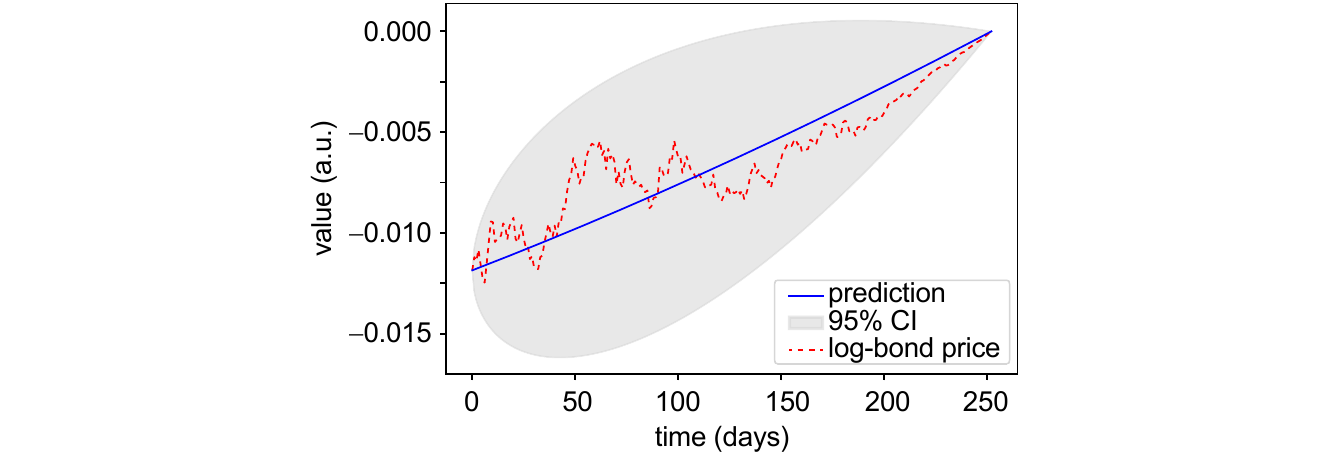}
  \caption{Illustration of a Gaussian process prior. Our task is to predict the unobserved log-bond-prices (red dotted line) with maturity $T=252$. The blue line represents the prior mean and the shaded area represents the 95\% confidence interval. Note that the initial value at time $0$ is known, as well as the final value $P(T,T)=1$ implying $\log P(T,T)=0$. }
  \label{fig:prior}
\end{figure}

For the dynamic prediction of  log-bond prices we seek a function that is able to predict new output values given some new input values, initially based on a set of training data. For this purpose we specify in a first step the {\em prior}\index{prior} distribution, expressing our prior beliefs over the function. In Figure~\ref{fig:prior} we plotted for the sake of illustration one simulated log-bond price and an arbitrary chosen prior, expressing our beliefs. This could correspond to the log-prices of a 1-year maturity bond, plotted over 252 days. The blue line constitutes the prior mean and the shaded area nearly twice the prior standard deviation corresponding to the 95\% confidence interval (CI). Without observing any training data, we are neither able to narrow the CI of our prior nor to enhance our prediction. 

\begin{figure}[t]
  \centering
  \includegraphics[width=\textwidth]{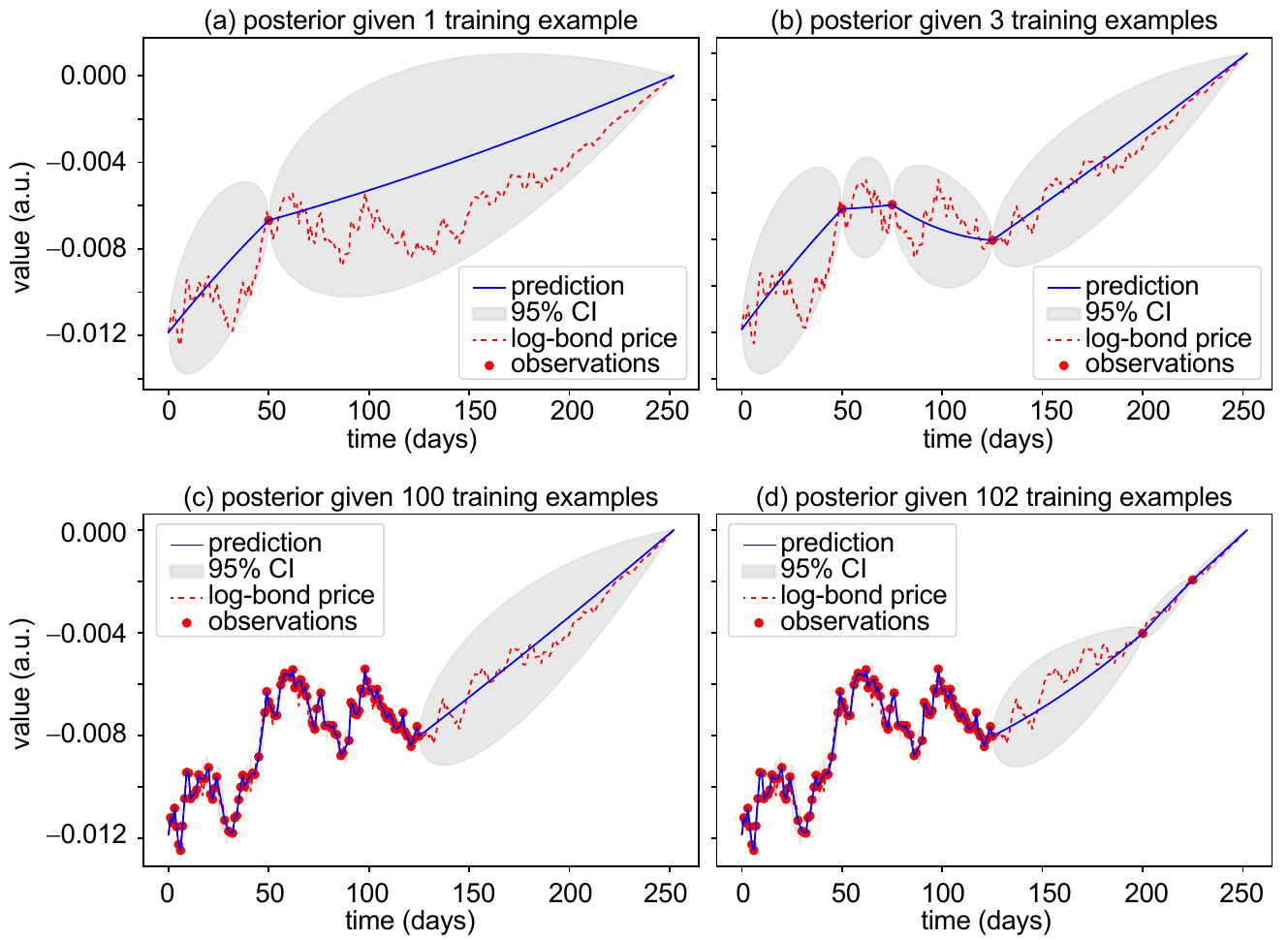}
  \caption{Illustration of Gaussian process posterior distributions, i.e. the prior conditioned on 1, 3, 100, and 102 observed training examples in (a), (b), (c), and (d), respectively. We want to predict log-bond-prices (red dotted line). The red dots constitute observations of the log-bond prices. The blue line denotes the posterior mean, which we take as prediction. The shaded areas represent the 95\% CIs. }
  \label{fig:post}
\end{figure}

The next step is to incorporate a set of training data, to enhance our prediction.
The Gaussian process regression has a quite fascinating property in this regard: consider an observation at the time point $t$, $t \in \{1,\dots,251\}$. The prediction for $y(t)$ is of course $y(t)$ itself - and hence perfect. The prediction of $y(s)$, be it for $s < t$ (which is often called \emph{smoothing}) or the prediction for $s>t$ (called \emph{filtering}) is normally distributed, but, according to \eqref{updating}, has a smaller variance compared to our initial prediction without the observation of $y(t)$. Increasing the number of observations, to, say, $y(t_1),\dots,y(t_n)$ improves the prediction dramatically. We illustrate this property in Figure \ref{fig:post}.
Depending on the location of $s$ relative to $t_1,\dots,t_n$ the prediction can become very exact (bottom right - for $s \in [0,130]$) and may still have a significant error (e.\,g. for $s$ around 180).

 We now formalize the question: assume we are
 given some  training data $y=(y(t_1,T),\dots,y(t_n,T))^\top$,  we aim at predicting the vector $\tilde y = (y(s_1,T),\dots,y(s_m,T))^\top$.

The joint distribution of $y$ and $\tilde y$
can directly be obtained from the mean and covariance functions in Equations \eqref{eq:mean0vas}, \eqref{eq:cov0vas}: since they are jointly normally distributed, we obtain that 
  \begin{align*}
  \left( 
  \begin{array}{l}
  y\\
  \tilde y
  \end{array}
  \right)  
  \sim \cN
  \left(
  \left(
  \begin{array}{l}
  \mu_y\\  
  \mu_{\tilde y}
  \end{array}
  \right)
  ,
  \left(
  \begin{array}{ll}
  \Sigma_{yy}
  &   \Sigma_{y \tilde y}\\
  \Sigma_{\tilde y y}
  &   \Sigma_{\tilde y \tilde y}        
  \end{array}
  \right)
  \right).
  \end{align*}
Recall that $\mu_y$ and $\Sigma_{yy}$ were already specified in Equation \eqref{distribution y}. Analogously, we obtain that 
\begin{align*}
  (\mu_{\tilde y})_i &= \mu(s_i,T) \\
  (\Sigma_{y \tilde y})_{i,j} &= c(t_i,s_j,T) \\
  (\Sigma_{\tilde y y})_{i,j} &= c(s_i,t_j,T) \\
  (\Sigma_{\tilde y \tilde y})_{i,j} &= c(s_i,s_j,T).
\end{align*}

The  {\em posterior} distribution  is obtained by 
using Equation \eqref{updating}: conditional on the training data $y$, we obtain that  
\begin{equation}\label{eq:posterior}
\tilde y | y 
\sim
\mathcal{N}\big(\tilde \mu,\tilde \Sigma\big),
\end{equation}
where
  \begin{equation*}
  \tilde \mu  = \mu_{\tilde y}
  +   \Sigma_{\tilde y y}
  \big(\Sigma_{yy} \big)^{-1}
  (y-\mu_y)
  \end{equation*}
and
  \begin{equation*}
  \tilde \Sigma
  = \Sigma_{\tilde y \tilde y}
  -  \Sigma_{\tilde y y}
  \big(\Sigma_{yy} \big)^{-1}
  \Sigma_{y \tilde y}.
  \end{equation*}
Given some training data, Equation \eqref{eq:posterior} now allows to calculate the posterior distribution, i.e. the conditional distribution of the predictions $\tilde y$ given the observation $y$,  and to make predictions, derive confidence intervals, etc. %

The posterior distributions, i.e. the prior conditioned on some training data of 1, 3, 100, and 102 observations, respectively, is shown in in Figure \ref{fig:post}.
The red dotted line represents a simulated trajectory of the log-bond price, while the red dots constitute the observed training data. 
The blue line is the posterior mean, which we take as our prediction, and the shaded areas denote the 95\% confidence interval (CI).  
We observe that the more training data we have, the more accurate our prediction gets. While in Figure~\ref{fig:post}~(a) the CI is quite large, two additional training data narrow the CI significantly down, see Figure~\ref{fig:post}~(b). Of course, the observation times $t_i$ play an important role in this: if we had observations $t_i$ close to $t_{i-1}$, the additional gain in knowledge would most likely be small, while in Figure ~\ref{fig:post}~(b), the $(t_i)$ are nicely spread. Consequently, if one is able to choose the $(t_i)$, one can apply results from optimal experimental design to achieve a highly efficient reduction in the prediction variance.

In most practical cases, the $(t_i)$ can not be chosen freely, and just arrive sequentially, as we illustrate in ~\ref{fig:post}~(c)--(d): here, we aim at  predicting log-bond prices of a 1-year maturity bond which we have observed daily until day $t_n=125$. Figure~\ref{fig:post}~(c) describes the situation in which we do not have any information about future prices of the 1-year maturity bond, i.e. the 2nd half of the 1 year. Figure~\ref{fig:post}~(d) depicts the situation in which we assume to know two log-bond prices or option strikes on zero-coupon bonds in the future, which enhances the prediction and narrows the CI down.

\subsection{Performance measures}

In order to monitor the performance of the calibration or the prediction,  one typically splits the observed data-set into two disjoint sets. The first set is called {\em training set} and is used to fit the model. The second one is the {\em validation set} and used as a proxy for the generalization error. Given the number of observed data points and number of parameters to optimize for our simulation task, we recommend to split the training and validation test set in a  $70\% / 30\%$- ratio. 
The quality of the predictions can be assessed  with the {\em standardized mean squared error} (SMSE) loss \index{standardized mean squared error} and the  {\em mean standardized log loss} (MSLL).\index{mean standardized log loss} 

The SMSE  considers the average of the squared residual between the mean prediction and the target of the test set and then standardizes it by the variance of the targets of the test cases. 
\begin{align*}
  \textup{SMSE} &= \frac{1}{\tilde{\sigma}^2}\frac{1}{m}\sum_{i =1}^{m} (\tilde y_i - \tilde \mu_i)^2,
\end{align*}
where $\tilde{\sigma}^2$ denotes the variance of the targets of the test cases.
The SMSE is a simple approach and the reader should bear in mind that this assessment of the quality does not incorporate information about the predictive variance. But as it represents a standard measure of the quality of an estimator, it is useful to consider it for the sake of comparability with other literature.

 The MSLL as defined in \cite{Rasmussen2006} is obtained by averaging the negative log probability of the target under the model over the test set and then standardizing it by subtracting the loss that would be obtained under the trivial model with mean and variance of the training data as in Equation~\eqref{eq:logmarglik}, 
\begin{align}\label{eq:MSLL}
  &\textup{MSLL} = -\frac{1}{m}\sum_{i=1}^{m} \log p(\tilde y_i| y ) + \log p(y) .
\end{align}
The MSLL will be around zero for simple methods and negative for better methods. Note that in \eqref{eq:MSLL} we omitted in the notation the dependence on the parameter set $\Theta$, since now the parameter set is fixed. Moreover, we notice that the MSLL incorporates the predictive mean and the predictive variances unlike the SMSE.

In the following section we simulate log-bond prices in a Vasi\v cek single-curve  model and calibrate the underlying parameters of the model. This is performed for one maturity. Since we deal with simulated prices, we consider the noise-free setting and set $\hat\sigma^2 = 0$ in  \eqref{eq:marglik}.

\subsection{Calibration results for the Vasi\v cek single-curve model}
We  generate 1.000 samples of log-bond price time series, each sample consisting of 250 consecutive prices (approximately one year of trading time) via Equations \eqref{eq:r} -- \eqref{eq:Vasicek_recall}. As parameter set we fix
$$ r_0 = 0.5, \quad \kappa = 2, \quad \theta = 0.1, \quad \sigma = 0.2. $$  %
For each of the 1.000 samples we seek the optimal choice of hyper-parameters  $\Theta = \{r_0, \kappa, \theta, \sigma\}$  based on the simulated trajectory of 250 log-bond prices. For this purpose we minimize the negative log marginal likelihood \eqref{eq:logmarglik} with underlying mean and covariance function \eqref{eq:mean0vas} and \eqref{eq:cov0vas}, respectively, by means of two optimization methods: the \emph{non-linear conjugate gradient} (CG)  algorithm and the \emph{adaptive moment estimation} (Adam) optimization algorithm. The CG algorithm uses a non-linear conjugate gradient by \cite{PolakRibiere1969}. For details on the CG algorithm we refer to \cite{nocedal2006conjugate}. The Adam \index{Adam} optimization algorithm is based on adaptive estimates of lower-order moments and is an extension of the stochastic gradient descent. For details of the Adam optimization algorithm we refer to \cite{kingma2014adam}.

 After a random initialization of the parameters to optimize, the CG optimization and the Adam optimization were performed using the python library {\em SciPy}\index{SciPy} and {\em TensorFlow}\index{TensorFlow}, respectively. Parallelization of the 1.000 independent runs was achieved with the python library {\em multiprocessing}. The outcomes can be summarized as follows.

\begin{table}[!t]
  
  \footnotesize
  \begin{center}
  \adjustbox{max height=\dimexpr\textheight-5.5cm\relax,
    max width=\textwidth}{
    \begin{tabular}{l | l l l l l l l l l}
      \hline
      \multirowcell{2}{\diagbox[height=2\lineskip,width =3cm]{\textbf{Optimizer}}{\textbf{Params.}}} 
      & \multicolumn{2}{c}{\textbf{$\boldsymbol{r_0}$}}   & \multicolumn{2}{c}{\textbf{$\boldsymbol{\kappa}$}}&  \multicolumn{2}{c}{\textbf{$\boldsymbol{\theta}$}}& 
      \multicolumn{2}{c}{$\boldsymbol{\sigma}$}&\\  
      & \textbf{Mean}& \textbf{StDev}&\textbf{Mean}& \textbf{StDev}&\textbf{Mean}& \textbf{StDev}&\textbf{Mean}& \textbf{StDev}\\
      \hline
      CG   & 0.496 & 0.135 &  2.081 & 0.474 & 0.104 & 0.106 & 0.202 & 0.020 \\
      Adam & 0.510   & 0.144 & 2.339 & 0.403 &  0.121 & 0.093  & 0.213 &0.018  \\
      \hline
      \hline
      \textbf{True value}   & 0.5 & & 2& &  0.1&  & 0.2   &\\
      \hline  
  \end{tabular}} \end{center}
  \caption{Single curve calibration results with mean and standard deviations (StDev) of the learned parameters (params.) of 1000 simulated log-bond prices as well as the true Vasi\v cek parameters.} 
  \label{table:sim_results_single}
\end{table}

\begin{figure}[!t]
  \centering
  \includegraphics[width=\textwidth]{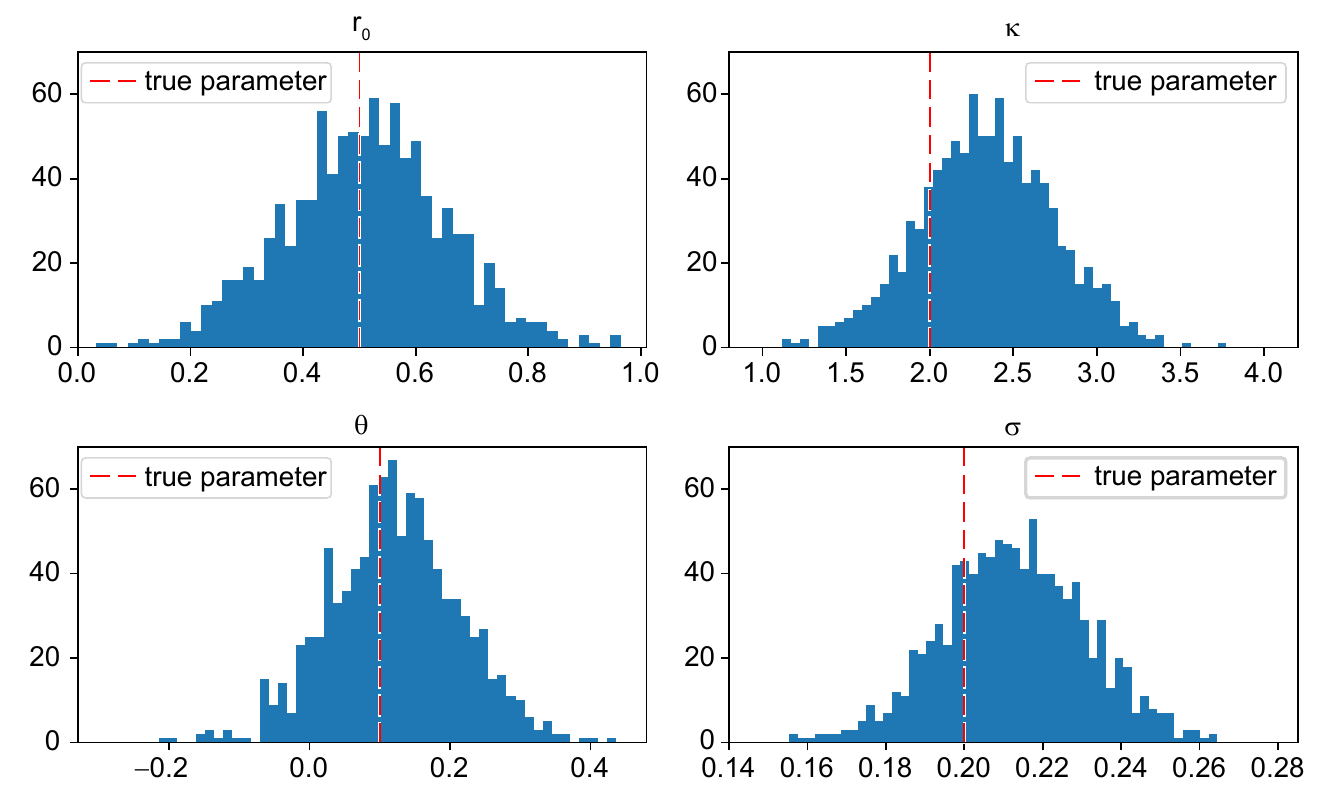}
  \caption{50 bins histogram of the learned parameters obtained with the Adam optimizer. The learning rate is 0.05 and training is performed over 700 epochs. Total simulations: 1000.}
  \label{fig:histAdam}
\end{figure}

\begin{enumerate}
\item  The results of the calibration via the conjugate gradient optimization algorithm are very satisfying. In Figure \ref{fig:histsingleCG} the learned parameters $r_0$, $\kappa$, $\theta$, and $\sigma$ of 1.000 simulated log-bond prices are plotted in 50 bins histograms. The red dashed lines in each sub-plot indicate the true model parameters.\\
 The mean and standard deviation of the learned parameters are summarized in Table~\ref{table:sim_results_single}. We observe that the mean of the learned parameters reaches the true parameters very closely while the standard deviation is reasonable.

\begin{figure}[!t]
  \centering
  \includegraphics[width=\textwidth]{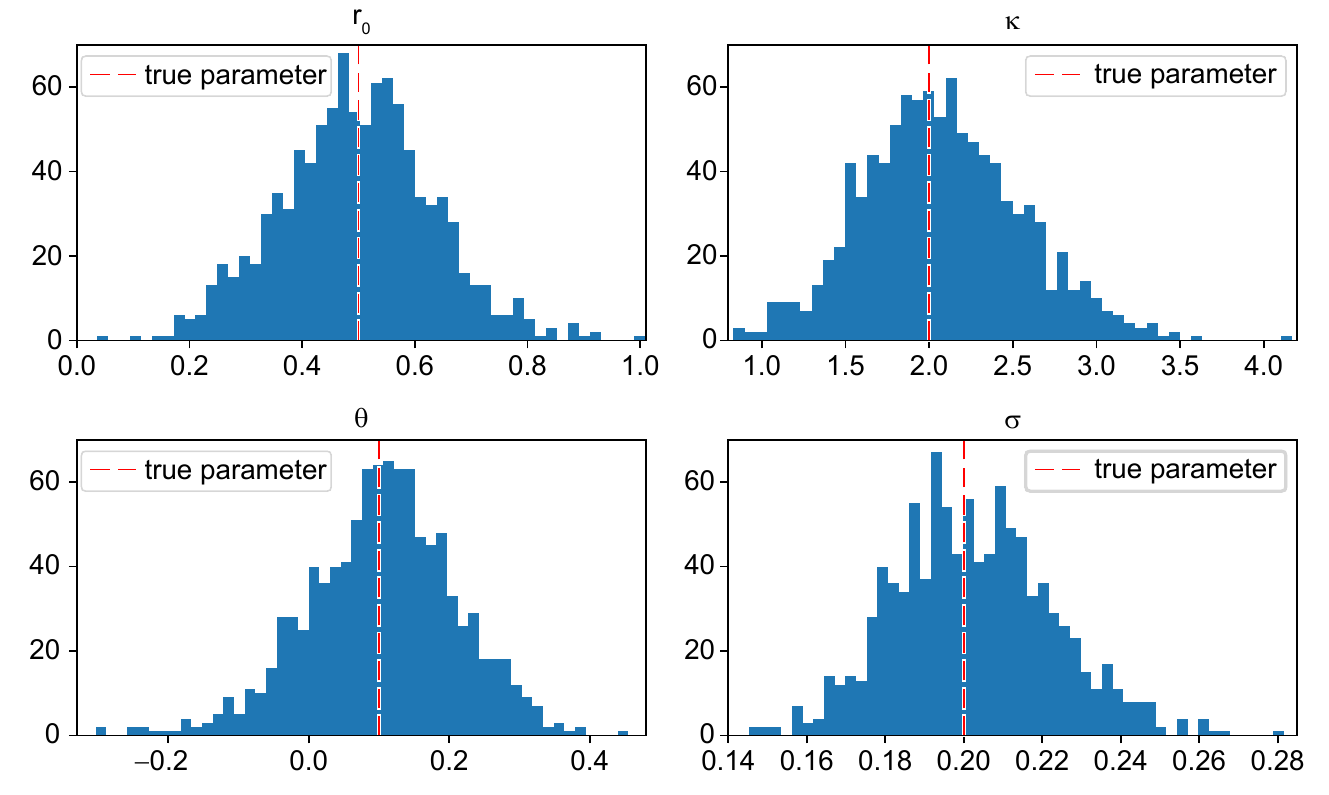}
  \caption{50 bins histogram of the learned parameters obtained with the CG optimizer. Total simulations: 1000.}
  \label{fig:histsingleCG}
\end{figure}

\item The results of the calibration via the Adam algorithm are satisfying, even though we note a shift of the mean reversion parameter $\kappa$ and the volatility parameter $\sigma$. The learned parameters $r_0$, $\kappa$, $\theta$, and $\sigma$ of 1000 simulated log-bond prices are plotted in 50 bins histograms in Figure \ref{fig:histAdam} and summarized with their mean and standard deviation in Table~\ref{table:sim_results_single}. The Adam algorithm slowly reduces the learning rate over time to speed up the learning algorithm. Nonetheless, one needs to specify a suitable learning rate. If the learning rate is small, training is more reliable, but the optimization time in order to find a minimum can increase rapidly. If the learning rate is too big, the optimizer can overshoot a minimum. We tried different learning rates of 0.0001, 0.001, 0.01, 0.05, and 0.1. Finally, we decided in favor of a learning rate of 0.05 and performed the training over 700 epochs in order to achieve a suitable trade-off between accuracy and computation time. We expect the results to improve slightly with more epochs, at the expense of a longer training time.
\end{enumerate}

We conclude that in the single-curve Vasi\v cek specification both optimization algorithms provide reliable results. The optimization by means of the CG algorithm outperforms the Adam algorithm, compare Table \ref{table:sim_results_single}.  However, there is hope that the results obtained by the Adam algorithm can be improved by increasing the number of training epochs or by choosing a smaller learning rate in addition to more training epochs.

\section{Multi-curve Vasi\v cek interest rate model}\label{subsec:multiVasicek}

The financial crisis in 2007--2009 has triggered many changes in financial markets.
In the post-crisis interest rate markets, multiple yield curves are standard: due to credit and liquidity issues, we obtain for each tenure a different curve, see for example \cite{GrbacRunggaldier} and the rich literature referenced therein.

The detailed mechanism in the multi-curve markets is quite involved, and we refer \cite{Grbacea2020} for a precise description. Intuitively, traded instruments are forward-rate agreements which exchange a fixed premium against a floating rate over a future time interval $[T,T+\delta]$. Most notably, in addition to the parameter maturity a second parameter appears: the tenor $\delta$. While before the crisis, the curves were independent of $\delta$, after the crisis the tenure can no longer be neglected. The authors in \cite{Grbacea2020} show that forward-rate agreements can be decomposed into $\delta$-bonds, which we denote by $P(t,T,\delta)$. If $\delta=0$ we write $P(t,T)=P(t,T,0)$, which is the single-curve case considered previously. 

In the multi-curve Vasi\v cek interest rate model we consider two maturity time points $T \leq T'$, hence  $\delta = T' -T$. We calibrate zero-coupon bond prices $P(t,T,0)$ and tenor-$\delta$-bond prices $P(t,T,\delta)$ under the risk-neutral measure $\Q$ at time $t$ for the maturity $T$.  

The calibration in the multi-curve framework corresponds to the situation in which we are given several data-sets  of different bond prices, where all of them are sharing the same hyper-parameters. We call this situation {\em multi-task learning} \index{multi-task learning} and reveal the different meaning in contrast to \cite[Section 5.4.3]{Rasmussen2006} where the corresponding log-marginal likelihoods of the individual problems are summed up and the result is then optimized with respect to the hyper-parameters. As our multi-curve example exhibits correlation, the latter approach can not be applied.

We will utilize a two-dimensional driving factor process $r=(r^1,r^2)^\top$, generalizing Equation \eqref{eq:r}. Modelling $r$ as a two-dimensional Ornstein--Uhlenbeck process can be done as follows: let $r$ be the unique solution of the SDE
\begin{equation}\label{eq:Vasicek_sde_recall}
dr^i_t = \kappa_i ( \theta_i - r^i_t) dt + \sigma_i dW^i_t, \quad t \ge 0, \ i=1,\ 2,
\end{equation}
where $W^1,W^2$ are two standard $\Q$-Brownian motions with correlation $\rho$. The zero-coupon bond prices only depend on $r^1$ and we utilize the single-curve affine framework:
\begin{equation}\label{eq:Vas_recall2}
P(t,T) = e^{-A(T-t)- B(T-t) r^1_t} \qquad \text{for } t \leq T.
\end{equation}
The functions $(A(\cdot),B(\cdot)):[0,T] \rightarrow \mathbb{R} \times \mathbb{R}$ satisfy the Riccati equations and we find in the new notation (replacing $\kappa$ by $\kappa_1$, etc.) that 
\begin{align*}
  \begin{split}
  A(T-t) = & \frac{\theta_1  \left(e^{-(T-t) \kappa_1 }-1+(T-t) \kappa_1 \right)}{\kappa_1 }\\
  &+\frac{\sigma_1^2}{4 \kappa_1^3}\big(e^{-2 (T-t) \kappa_1 }-4 e^{-(T-t) \kappa_1 }-2
    (T-t) \kappa_1 +3\big) \\
  B(T-t) = & \frac{1}{\kappa_1 }\big(1 -e^{-\kappa_1(T-t)  }\big).
  \end{split}
\end{align*}
As in \eqref{eq:mean0vas} and \eqref{eq:cov0vas} we obtain that zero-coupon log-bond prices $y(t,T,0)=\log P(t,T,0)$ are normal with mean function
\begin{align}\label{eq:mean0vasmulti}
\mu(t,T,0)& := \E_{\Q}\big[\log P(t,T,0)\big]\nonumber\\
&= -A(T-t)-B(T-t)\big(r^1_0 e^{-\kappa_1 t} + \theta_1 (1-e^{-\kappa_1 t})\big),\quad t\leq T
\end{align}
For two given time points $t_i,t_j \leq T$ the covariance function is 
\begin{align}\label{eq:cov0vasmulti}
c(t_i,t_j,T,0) &:=
 B(T -t_i)B(T-t_j) \frac{\sigma_1^2}{2\kappa_1} e^{-\kappa_1(t_i +t_j)} \big(e^{2\kappa_1 (t_i \wedge t_j)} -1\big).
\end{align}

The next step is to develop the prices for the tenor-$\delta$ bonds. We assume that while for the tenor $0$ the interest rate is $r^1$, the interest rate for tenor $\delta$ is $r^1 + r^2$. This implies that
$$  P(t,T,\delta)
 = \E_{\Q}\Big[e^{-\int_t^T (r^1_s - r^2_s) ds} \Big| \mathcal{F}_t\Big]. $$
Since $r$ is an affine process, this expectation can be computed explicitly. Using the affine machinery we obtain that 
\begin{align}
P(t,T,\delta)
& = 
 \exp\big({\Phi(T-t) + \Psi(T-t)^\top r_t}\big), \qquad \text{for }t \leq T \label{eq:deltaBondreprexplrecall}
\end{align}
where the function $(\Phi(\cdot),\Psi(\cdot)):[0,T] \rightarrow \mathbb{R} \times \mathbb{R}^2$ satisfies the Riccati equations. This implies that
\begin{align}\label{eq:verFilipmultirecall}
\Psi_1(T-t) &= - B(T-t) = -\frac{1}{\kappa_1}\big(1-e^{-\kappa_1(T-t)}\big), \nonumber \\
\Psi_2(T-t) &=   \frac{1}{\kappa_2}\big(1-e^{-\kappa_2(T-t)}\big) \text{ and}\nonumber\\
\Phi(T-t) %
&= - (\theta_1-\theta_2)(T-t)- \frac{\theta_1}{\kappa_1}\big(e^{-\kappa_1(T-t)}-1\big) 
+ \frac{\theta_2}{\kappa_2}\big(e^{-\kappa_2(T-t)}-1\big)\nonumber\\
&\quad+\frac{\sigma_1^2}{2\kappa_1^2}\Big(T-t + \frac{2}{\kappa_1}e^{-\kappa_1(T-t)}-\frac{1}{2\kappa_1}e^{-2\kappa_1(T-t)}-\frac{3}{2\kappa_1}\Big)\nonumber\\
&\quad +\frac{\sigma_2^2}{2\kappa_2^2}\Big(T-t + \frac{2}{\kappa_2}e^{-\kappa_2(T-t)}-\frac{1}{2\kappa_2}e^{-2\kappa_2(T-t)}-\frac{3}{2\kappa_2}\Big)\nonumber\\
&\quad - \frac{\rho\sigma_1\sigma_2}{\kappa_1\kappa_2}\Big(T-t + \frac{1}{\kappa_1}(e^{-\kappa_1(T-t)}-1)
+ \frac{1}{\kappa_2}(e^{-\kappa_2(T-t)}-1)\nonumber\\
&\quad-\frac{1}{\kappa_1+\kappa_2}\big(e^{-(\kappa_1+\kappa_2)(T-t)}-1\big)\Big).
\end{align}

\begin{proposition} \label{prop1}
Under the above assumptions, 
 $\log P(\cdot,T,\delta)$, $0 \le t \le T$ is a Gaussian process with mean function 
\begin{align*}
  \mu(t,T,\delta)
  & = \Phi(T-t) 
  + \Psi_1(T-t)(r_0^1 e^{-\kappa_1 t} + \theta_1 (1-e^{-\kappa_1 t}))\\
  &\quad
  + \Psi_2(T-t)(r_0^2 e^{-\kappa_2 t} + \theta_2 (1-e^{-\kappa_2 t})), \quad t \leq T,
\end{align*} 
and with covariance function 
\begin{align*}
  c(s,t,T,\delta)
  &=  \Psi_1(T-s) \Psi_1(T-t)\frac{\sigma_1^2}{2\kappa_1} e^{-\kappa_1(s +t)} \big(e^{2\kappa_1 (s \wedge t)} -1\big)\\
  &\quad+ \Psi_1(T-s) \Psi_2(T-t)\frac{ \rho \sigma_1\sigma_2}{\kappa_1 + \kappa_2}  e^{-(\kappa_1 s+\kappa_2 t)}\big(e^{(\kappa_1 + \kappa_2)(s \wedge t)} -1\big)\\
  &\quad+ \Psi_1(T-t) \Psi_2(T-s)\frac{ \rho \sigma_1\sigma_2}{\kappa_1 + \kappa_2}  e^{-(\kappa_1 t+\kappa_2 s)}\big(e^{(\kappa_1 + \kappa_2)(t \wedge s)} -1\big)\\
  &\quad+ \Psi_2(T-s) \Psi_2(T-t)\frac{\sigma_2^2}{2\kappa_2} e^{-\kappa_2(s +t)} \big(e^{2\kappa_2 (s \wedge t)} -1\big), \quad 0 \le s,t \le T.
\end{align*}
\end{proposition}
Note that $\mu$ and $c$ in the above proposition do not depend on $\delta$. We rather use $\delta$ as index to distinguish mean and covariance functions for the zero-coupon bonds and the $\delta$-tenor bonds.

The proof is relegated to the appendix. 
The next step is to phrase observation and prediction in the Gaussian setting explicitly. 
We denote $y=(y^0,y^\delta)^\top$ where 
$y^0 = (y(t_1,T,0),\dots,y(t_n,T,0))$ and $y^\delta = (y(t_1,T,\delta),\dots,y(t_n,T,\delta))$ -- of course, at each time point $t_i$ we observe \emph{two} bond prices now: $P(t_i,T,0)$ and $P(t_i,T,\delta)$. The vector $y$ is normally distributed, $y \sim \ccN(\mu_y,\Sigma_y)$ with 
\begin{align}\label{eq:muvasmulti}
  \mu_y
  := 
  \left(
  \begin{array}{l}
  \mu(\cdot, T, 0)\\   
  \mu(\cdot, T, \delta)
  \end{array}
  \right), \qquad 
\Sigma_y 
    :=
    \left(
    \begin{array}{ll}
    \Sigma_y^{00}
    &  \Sigma_y^{0\delta}\\
    \Sigma_y^{\delta 0}
    &   \Sigma_y^{\delta \delta}          
    \end{array}
    \right).
\end{align}
We calculate these parameters explicitly and, analogously to the single-curve set-up, the calibration methodology with Gaussian process regression follows.

To begin with, note that $\Sigma_y^{00}$ coincides with $\Sigma_{yy}$ from equation \eqref{distribution y}, when we replace the parameters $r_0,\kappa,\theta,\sigma$ by $r_0^1,\kappa^1,\theta^1,\sigma^1$, respectively. The next step for computing the covariance matrix is to compute
\begin{align*}
  &(\Sigma_y^{0 \delta})_{i,j}\\
  &:= \E_\Q \Big[\big(\log P(t_i,T,0) - \E_\Q\big[\log P(t_i,T,0)\big]\big)
  \cdot \big(\log P(t_j,T,\delta) - \E_\Q\big[\log P(t_j,T,\delta)\big]\big)\Big] \\
  &=  \Psi_1(T -t_i)\Psi_1(T-t_j) \frac{\sigma_1^2}{2\kappa_1} e^{-\kappa_1(t_i +t_j)} (e^{2\kappa_1 (t_i \wedge t_j)} -1)\\
  &\quad+\,\Psi_1(T-t_i) \Psi_2(T-t_j)\frac{ \rho \sigma_1\sigma_2}{\kappa_1 + \kappa_2}  e^{-(\kappa_1 t_i+\kappa_2 t_j)}\big(e^{(\kappa_1 + \kappa_2)(t_i \wedge t_j)} -1\big),
\end{align*}
and, analogously,
\begin{align*}
&(\Sigma_y^{\delta 0})_{i,j}\\
&:= \E_\Q \Big[ \big(\log P(t_i,T,\delta) - \E_\Q\big[\log P(t_i,T,\delta)\big]\big)
 \cdot\big(\log P(t_j,T,0) - \E_\Q\big[\log P(t_j,T,0)\big]\big)\Big] \\
&=  \Psi_1(T -t_i)\Psi_1(T-t_j) \frac{\sigma_1^2}{2\kappa_1} e^{-\kappa_1(t_i +t_j)} (e^{2\kappa_1 (t_i \wedge t_j)} -1)\\
&\quad+\,\Psi_1(T-t_j) \Psi_2(T-t_i)\frac{ \rho \sigma_1\sigma_2}{\kappa_1 + \kappa_2}  e^{-(\kappa_1 t_j+\kappa_2 t_i)}\big(e^{(\kappa_1 + \kappa_2)(t_i \wedge t_j)} -1\big);
\end{align*}
in a similar way to $\Sigma_y^{00}$ we obtain $\Sigma_y^{\delta \delta}$.

In the multi-curve calibration we aim at minimizing the log marginal likelihood \eqref{eq:logmarglik} with corresponding mean function  and covariance matrix \eqref{eq:muvasmulti}.

\subsection{Calibration results}
For the calibration in the multi-curve setting we have to consider log-bond prices and the logarithm of tenor-$\delta$ bond prices. To be specific, we generate 1000 sequences of log-bond prices and 1000 sequences of log tenor-$\delta$ bond prices with maturity 1  by means of \eqref{eq:Vas_recall2}, and \eqref{eq:deltaBondreprexplrecall}--\eqref{eq:verFilipmultirecall}. We choose as parameters
\begin{align*}
  r^1_0 &= 0.5,\  \kappa_1 = 2, \quad \theta_1 = 0.1, \quad \sigma_1 = 0.2 \\
  r^2_0 &= 0.7, \ \kappa_2 = 0.5, \ \theta_2 = 0.03, \ \ \sigma_2 = 0.8.
\end{align*} 
 For each sequence we generate 125 training data points of log-bond prices and 125 training data points of log tenor-$\delta$ bond prices.  This corresponds to a computational effort similar to the single-curve specifications, since the underlying covariance matrix will be $(250\times250)$-dimensional. Based on the simulated prices we aim at finding the optimal choice of \emph{hyper-parameters} 
 $$ \Theta = \{r^1_0, \kappa_1, \theta_1 , \sigma_1, r^2_0,\kappa_2, \theta_2, \sigma_2\}. $$ For this purpose, we apply the CG optimization algorithm and the Adam optimization algorithm. After a random initialization of the parameters to optimize, we perform the calibration procedure using SciPy and TensorFlow for the CG and Adam optimizer, respectively. Parallelization of the 1000 independent runs is achieved with the library multiprocessing. The outcomes are as follows.
\begin{table}[!t]
  \begin{center}
    \adjustbox{max height=\dimexpr\textheight-5.5cm\relax,
    max width=\textwidth}{\begin{tabular}{l | l l l l l l l l l}
      \hline
      \multirowcell{2}{\diagbox[height=2\lineskip, width =3cm]{\textbf{Optimizer}}{\textbf{Params.}}} 
      & \multicolumn{2}{c}{\textbf{$\boldsymbol{r^1_0}$}}   & \multicolumn{2}{c}{\textbf{$\boldsymbol{\kappa_1}$}}&  \multicolumn{2}{c}{\textbf{$\boldsymbol{\theta_1}$}}& 
      \multicolumn{2}{c}{$\boldsymbol{\sigma_1}$}&\\  
      & \textbf{Mean}& \textbf{StDev}&\textbf{Mean}& \textbf{StDev}&\textbf{Mean}& \textbf{StDev}&\textbf{Mean}& \textbf{StDev}\\
      \hline
      CG & 0.477& 0.119 & 1.994 &  0.781  & 0.101 & 0.121 & 0.150 & 0.033 \\     
      Adam & 0.487& 0.053& 2.694 & 0.284  & 0.157 & 0.029  & 0.170 &0.013\\
      \hline
      \hline
      \textbf{True value}   & 0.5 & & 2& &  0.1&  & 0.2   &\\
      \hline  
  \end{tabular}}
  \adjustbox{max height=\dimexpr\textheight-5.5cm\relax,
    max width=\textwidth}{
    \begin{tabular}{l | l l l l l l l l l}
      \hline
      \multirowcell{2}{\diagbox[height=2\lineskip, width =3cm]{\textbf{Optimizer}}{\textbf{Params.}}} 
      & \multicolumn{2}{c}{\textbf{$\boldsymbol{r^2_0}$}}   & \multicolumn{2}{c}{\textbf{$\boldsymbol{\kappa_2}$}}&  \multicolumn{2}{c}{\textbf{$\boldsymbol{\theta_2}$}}& 
      \multicolumn{2}{c}{$\boldsymbol{\sigma_2}$}&\\  
      & \textbf{Mean}& \textbf{StDev}&\textbf{Mean}& \textbf{StDev}&\textbf{Mean}& \textbf{StDev}&\textbf{Mean}& \textbf{StDev}\\
      \hline
      CG & 0.678 & 0.601 & 0.529 & 0.545 & 0.385 & 1.732 & 0.602  &  0.113\\
      Adam  & 0.477 & 0.139 & 0.813  & 0.277 & 0.876 & 0.254 & 0.640 & 0.055\\
      \hline
      \hline
      \textbf{True value}   & 0.7 & & 0.5& &  0.03&   & 0.8   &\\
      \hline  
  \end{tabular}} \end{center} 
    \caption{Calibration results with mean and standard deviations (StDev) of the learned parameters (params.) of 1000 simulated log-bond prices and log tenor-$\delta$ bond prices, as well as the true Vasi\v cek parameters. We note that the CG optimizer yields for every parameter a higher standard deviation than the Adam algorithm. This can in particular be observed for the parameter $\theta_2$. } \label{table:sim_results_multi}
\end{table}

\begin{figure}[!ht]
	\centering
	\includegraphics[width=\textwidth]{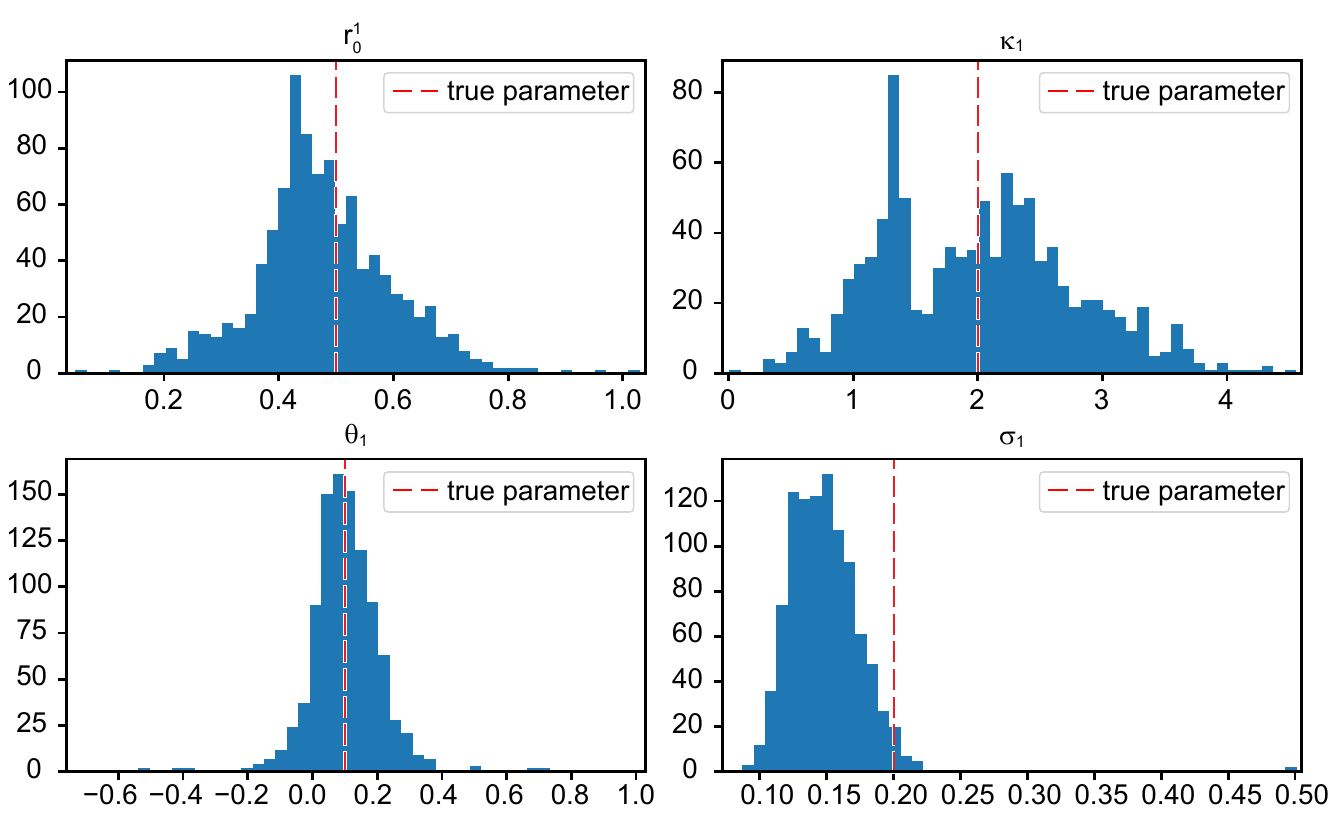}
	\includegraphics[width=\textwidth]{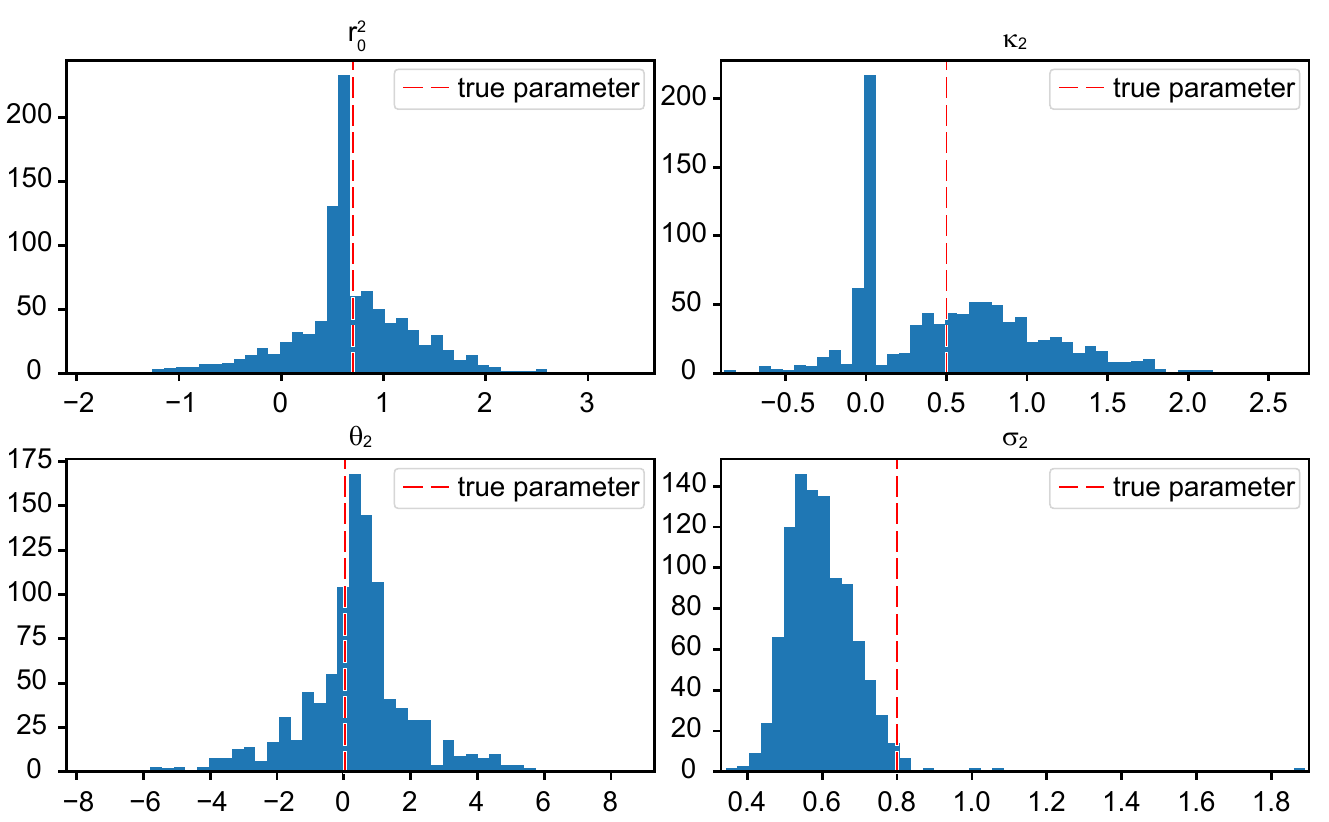}
	\caption{50 bins histogram of the learned parameters with the conjugate gradient (CG) optimizer. Total simulations: 1000. 
    We note, that the learned parameters are centered around the true parameter values except for the volatility parameters $\sigma_1$ and $\sigma_2$, exhibiting a shift. In statistics, the task of estimating the volatility parameters is well known. However, our observations reveal, that the task of finding the volatility parameter in the multi-curve Vasi\v cek interest rate model constitutes a challenging task. Moreover, the range of the parameters, especially for $\theta_2$, is very high.}
	\label{fig:hist_multi_CG_1000}
\end{figure}

\begin{figure}[!ht]
	\centering
	\includegraphics[width=\textwidth]{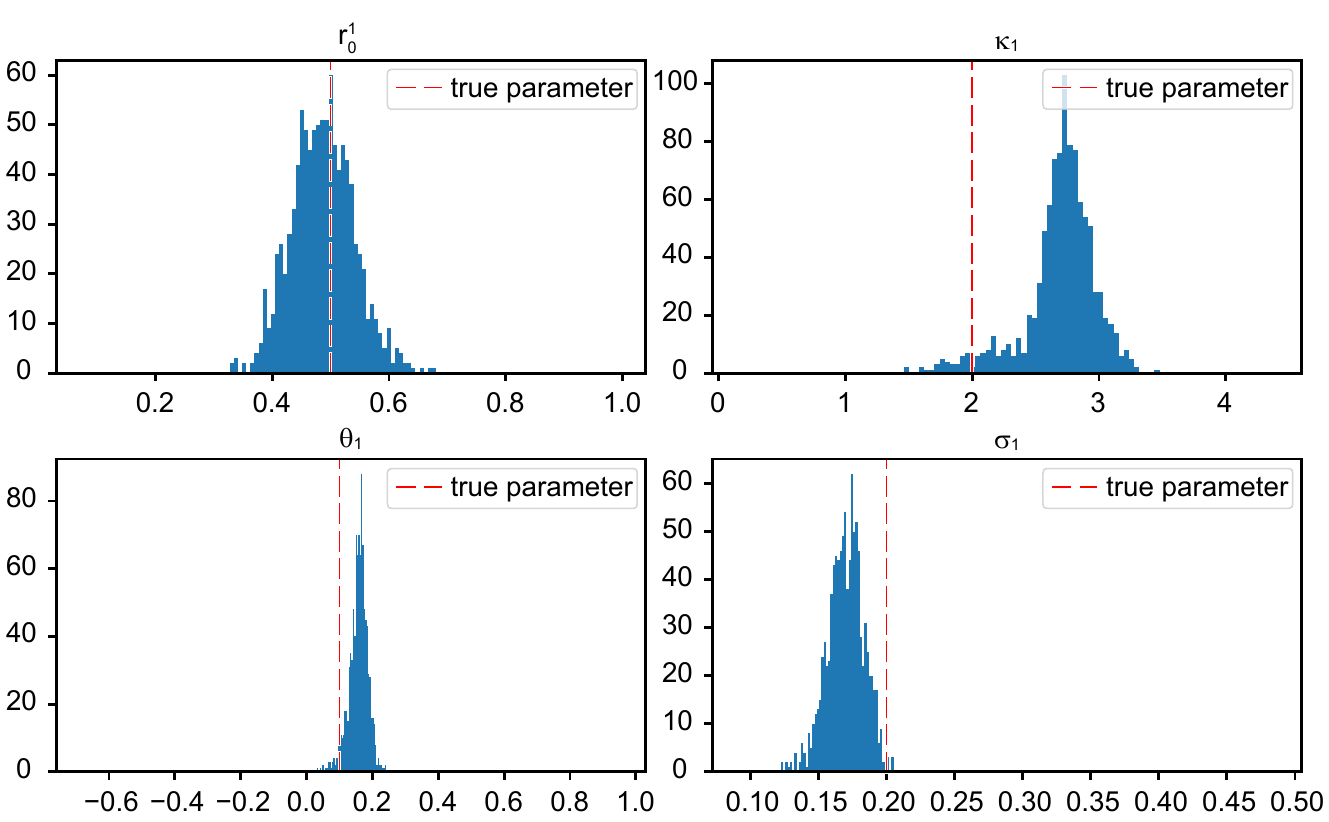}
	\includegraphics[width=\textwidth]{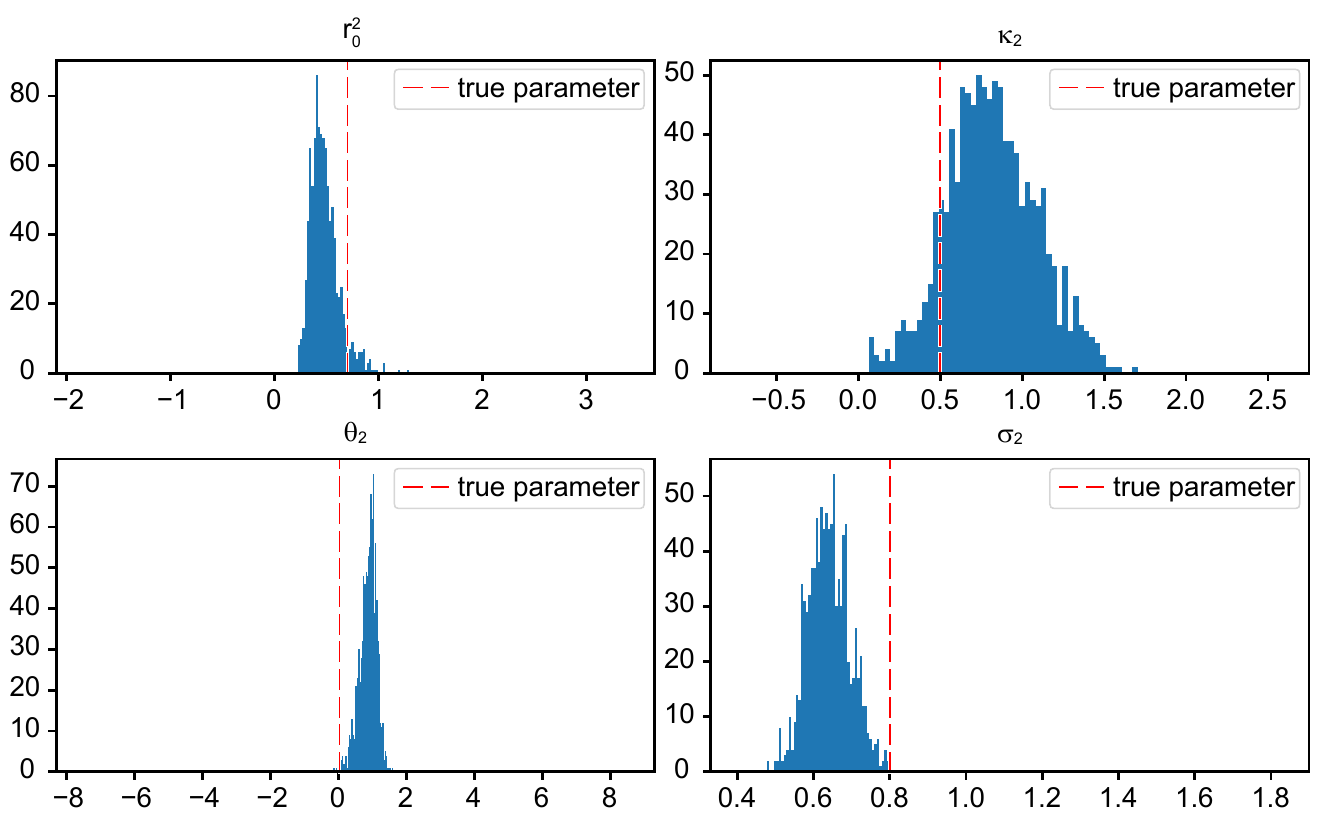}
	\caption{50 bins histogram of the learned parameters obtained with the adaptive moment estimation (Adam) optimizer. The learning rate is 0.05 and training is performed over 750 epochs. Total simulations: 1000. We note, that the range of learned parameters is more narrow compared to the CG optimizer, resulting in a lower standard deviation, cf. Table \ref{table:sim_results_multi}.}
	\label{fig:hist_multi_Adam_1000}
\end{figure}

\begin{enumerate}
  \item Considering the results of the calibration by means of the CG algorithm, we note several facts. In this regard, the mean and the standard deviation of the calibrated parameters can be found in Table~\ref{table:sim_results_multi}.   Figure~\ref{fig:hist_multi_CG_1000} shows the learned parameters of the processes $r^1$ and $r^2$ in 50 bins histograms for 1000 simulations. The red dashed line in each of the sub-plots indicates the true model parameter value. 
  First, except for the long term mean, the volatility for $r^2$ is higher than the one for $r^1$ which implies more difficulties in the estimation of the parameters for $r^2$, which is clearly visible in the results. For $r^1$, we are able to estimate the parameters well (in the mean), with the most difficulty in the estimation of $\kappa^1$ which shows a high standard deviation.

  For estimating the parameters of $r^2$ we face more difficulties, as expected. The standard deviation of $\theta_2$ is very high -- it is known from filtering theory and statistics that the mean is difficult to estimate, which is reflected here. Similarly, it seems difficult to estimate the speed of reversion parameter $\kappa_2$ and we observe a peak around 0.02 in $\kappa_2$. This might be due to a local minimum, where the optimizer gets stuck.

  \item For the calibration results by means of the Adam algorithm we note the following. The learned parameters are illustrated in a 50 bins histogram, see Figure~\ref{fig:hist_multi_Adam_1000}, and  mean and standard deviation of each parameter are stated in Table~\ref{table:sim_results_multi}. After trying several learning rates of 0.0001, 0.001, 0.01, 0.05, and 0.1 we decided in favor of the learning rate 0.05 and chose 750 training epochs. While most mean values of the learned parameters are not as close to the true values as the learned parameters from the CG algorithm, we notice, that the standard deviation of the learned parameters is smaller compared to the standard deviation of the learned parameters  from the CG. Especially, comparing the values of $\theta_2$ in Figure \ref{fig:hist_multi_CG_1000} and Figure \ref{fig:hist_multi_Adam_1000}, we observe the different range of calibrated values.
\end{enumerate}

\section{Conclusion}

Concluding the simulation results, we can state that the calibration in the multi-curve framework constitutes a more challenging task compared to the calibration in the single-curve framework, since we need to find 8 parameters instead of 4 parameters. In particular, in order to keep the computational complexity similar to the single-curve framework, we chose 125 training data time points, which results in a covariance matrix of the same dimension as in the single-curve setting. We are confident that doubling the training input data points would improve the results at the expense of computation time.

It would also be interesting to analyze, how other estimators perform in comparison to the shown results, like for example classical maximum-likelihood estimators. Since it is already known that it is difficult to estimate the variance with ML-techniques, it could also be very interesting to mix classical approaches with ML approaches, which we leave for future work.

A first step in order to extend the presented approach could consist in studying further optimization techniques such as the adaptive gradient algorithm (AdaGrad) or its extension Adadelta, the root mean square propagation (\hbox{RMSProp}), and the Nesterov accelerated gradient (NAG), one could further investigate the prediction of log-bond prices with the learned parameters in order to obtain decision-making support for the purchase of options on zero-coupon bonds and use the underlying strike prices as worst-case scenarios. A next step could be the development of further short rate models such as the Cox--Ingersoll--Ross, the Hull--White extended Vasi\v cek or  Cox--Ingersoll--Ross framework. An interesting application is the calibration of interest rate models including jumps. Beyond that, the study of the calibration of interest rate markets by means of Bayesian neural networks seems very promising and remains to be addressed in future work.

The Gaussian process regression approach naturally comes with the a posteriori distribution, which contains much more information compared to the simple prediction (which contains only the mean). It seems to be highly interesting to utilize this for assessing the model risk of the calibration and compare it to the non-linear approaches recently developed in \cite{FadinaNeufeldSchmidt2019,holzermann2020pricing}.

Summarizing, the calibration of multiple yield curves is a difficult task and we hope to stimulate future research with this initial study showing promising results on the one side and many future challenges on the other side.

\begin{appendix}

\section{Proof of Proposition \ref{prop1}}

First, we calculate the mixed covariance function: for two time points $t_i,t_j \leq T$ we obtain that 
\begin{align*}
  \E_\Q[r^1_{t_i} r^2_{t_j}] 
  &= \E_\Q\Big[\big(r^1_0 e^{-\kappa_1 t_i} + \theta_1(1-e^{-\kappa_1 t_i}) + \sigma_1 e^{-\kappa_1 t_i} \int_0^{t_i} e^{\kappa_1 u} dW^1_u \big)\\
  &\quad \quad \cdot\big(r^2_0 e^{-\kappa_2 t_j} + \theta_2(1-e^{-\kappa_2 t_j}) + \sigma_2 e^{-\kappa_2 t_j} \int_0^{t_j} e^{\kappa_2 u} dW^2_u \big)\Big]\\
  & = r^1_0 r^2_0 e^{-(\kappa_1 t_i+\kappa_2 t_j)} + r^1_0 \theta_2e^{-\kappa_1 t_i} (1-e^{-\kappa_2 t_j})
   \\
  & \quad+ r^2_0 \theta_1 e^{-\kappa_2 t_j}  (1-e^{-\kappa_1 t_i}) +  \theta_1 \theta_2(1-e^{-\kappa_1 t_i}) (1-e^{-\kappa_2 t_j})\\
  &
  \quad+  \sigma_1\sigma_2 e^{-(\kappa_1 t_i+\kappa_2 t_j)}\E_\Q\Big[\int_0^{t_i} e^{\kappa_1 u} dW^1_u  \int_0^{t_j} e^{\kappa_2 u} dW^2_u\Big]\\
  &= r^1_0 r^2_0 e^{-(\kappa_1 t_i+\kappa_2 t_j)} + r^1_0 \theta_2e^{-\kappa_1 t_i} (1-e^{-\kappa_2 t_j})
  + r^2_0 \theta_1 e^{-\kappa_2 t_j}  (1-e^{-\kappa_1 t_i}) \\
  & \quad+  \theta_1 \theta_2(1-e^{-\kappa_1 t_i}) (1-e^{-\kappa_2 t_j})
  + \rho \sigma_1\sigma_2 e^{-(\kappa_1 t_i+\kappa_2 t_j)}\int_0^{t_i \wedge t_j } e^{(\kappa_1 + \kappa_2) u} du\\
  &= r^1_0 r^2_0 e^{-(\kappa_1 t_i+\kappa_2 t_j)} + r^1_0 \theta_2e^{-\kappa_1 t_i} (1-e^{-\kappa_2 t_j}) \\
  &  \quad+ r^2_0 \theta_1 e^{-\kappa_2 t_j}  (1-e^{-\kappa_1 t_i}) +  \theta_1 \theta_2(1-e^{-\kappa_1 t_i}) (1-e^{-\kappa_2 t_j})\\
  &
  \quad+\frac{ \rho \sigma_1\sigma_2}{\kappa_1 + \kappa_2}  e^{-(\kappa_1 t_i+\kappa_2 t_j)}(e^{(\kappa_1 + \kappa_2)(t_i \wedge t_j)} -1).
\end{align*}
For the computation of $\E_\Q[\int_0^{t_i} e^{\kappa_1 u} dW^1_u  \int_0^{t_j} e^{\kappa_2 u} dW^2_u]$ we applied \cite[Theorem I-4.2]{JacodShiryaev}. Furthermore, note  that for two Brownian motions $W^1,W^2$ with correlation $\rho$, we find a Brownian motion $W^3$, independent of $W^1$, such that $W^2 = \rho W^1 + \sqrt{(1-\rho^2)} W^3$ and  that a Brownian motion has independent increments.

Hence, we obtain that the tenor-$\delta$-bond prices are log-normally distributed with mean function   
\begin{align*}
  \mu(t,T,\delta)&= \E_{\Q}[\log P(t,T,\delta)]  = \Phi(T-t) 
  + \Psi_1(T-t)(r_0^1 e^{-\kappa_1 t} + \theta_1 (1-e^{-\kappa_1 t}))\\
  &
  \quad+ \Psi_2(T-t)(r_0^2 e^{-\kappa_2 t} + \theta_2 (1-e^{-\kappa_2 t})), \quad t \leq T,
\end{align*} 
and with covariance function for $t_i,t_j \leq T$
\begin{align*}
  c&(t_i,t_j,T,\delta)\\
  &=
  \E_{\Q}\Big[\big(\log P(t_i,T,\delta) - \E_{\Q}[\log P(t_i,T,\delta)]\big)
  \cdot \big(\log P(t_j,T,\delta) - \E_{\Q}[\log P(t_j,T,\delta)]\big)\Big]\\
  &=  \Psi_1(T-t_i) \Psi_1(T-t_j)\big(\E_\Q[ r^1_{t_i} r^1_{t_j}] - \E_\Q[ r^1_{t_i}]\E_\Q [r^1_{t_j}]\big)\\
  & \quad+\Psi_1(T-t_i) \Psi_2(T-t_j)\big(\E_\Q[ r^1_{t_i} r^2_{t_j}] - \E_\Q[ r^1_{t_i}]\E_\Q [r^2_{t_j}]\big)\\
  & \quad+ \Psi_1(T-t_j) \Psi_2(T-t_i)\big(\E_\Q[ r^1_{t_j} r^2_{t_i}] - \E_\Q[ r^1_{t_j}]\E_\Q [r^2_{t_i}]\big) \\
  &\quad +\Psi_2(T-t_i) \Psi_2(T-t_j)\big(\E_\Q[ r^2_{t_i} r^2_{t_j}] - \E_\Q[ r^2_{t_i}]\E_\Q [r^2_{t_j}]\big)\\
  &=  \Psi_1(T-t_i) \Psi_1(T-t_j)\frac{\sigma_1^2}{2\kappa_1} e^{-\kappa_1(t_i +t_j)} \big(e^{2\kappa_1 (t_i \wedge t_j)} -1\big)\\
  &\quad+ \Psi_1(T-t_i) \Psi_2(T-t_j)\frac{ \rho \sigma_1\sigma_2}{\kappa_1 + \kappa_2}  e^{-(\kappa_1 t_i+\kappa_2 t_j)}\big(e^{(\kappa_1 + \kappa_2)(t_i \wedge t_j)} -1\big)\\
  &\quad+ \Psi_1(T-t_j) \Psi_2(T-t_i)\frac{ \rho \sigma_1\sigma_2}{\kappa_1 + \kappa_2}  e^{-(\kappa_1 t_j+\kappa_2 t_i)}\big(e^{(\kappa_1 + \kappa_2)(t_j \wedge t_i)} -1\big)\\
  &\quad+ \Psi_2(T-t_i) \Psi_2(T-t_j)\frac{\sigma_2^2}{2\kappa_2} e^{-\kappa_2(t_i +t_j)} \big(e^{2\kappa_2 (t_i \wedge t_j)} -1\big).
\end{align*}

\section{Coding notes and concluding remarks}\label{sec:codingconcluding}
We add some valuable remarks regarding the coding framework and deliver insight to the technical difficulties that arose during the calibration exercise.

In our application we needed customized covariance matrices. Therefore, we decided against applying one of the Python libraries for the implementation of Gaussian process regression (amongst others the packages {\em pyGP}, {\em pyGPs}, {\em scikit-learn}, {\em GPy}, {\em Gpytorch}, {\em GPflow}) due to their limited choice of covariance functions. We aimed at achieving a trade-off between computational speed and accuracy. We performed all simulations in two environments. The first comprises an AMD\textsuperscript{\textregistered} Ryzen 2700x CPU equipped with 16 cores, and one GeForce GTX 1070 GPU. The second comprises 4 Intel(R) Xeon(R) Gold 6134 CPU resulting in 32 cores, and 4 GeForce GTX 1080 GPUs. Since the calibration of the different settings was performed in two environments with different hardware specifications, we do not compare the optimizers regarding time consumption.

Utilizing all available resources comprising CPUs and GPUs for parallel operations also turned out to be  a delicate task. For example the TensorFlow default distribution is built without extensions such as AVX2 and FMA. Our CPU supports the advanced vector extensions AVX2 FMA, requiring  the build from source of the library TensorFlow. The reader should bear in mind that also minor code improvements of functions which are called the most, improve the overall performance considerably.

\end{appendix}

%


%
\end{document}